\documentclass[11pt]{article}

\usepackage[margin=1.1in]{geometry}
\usepackage[T1]{fontenc}
\usepackage[utf8]{inputenc}
\usepackage{lmodern}
\IfFileExists{microtype.sty}{\usepackage{microtype}}{}
\usepackage{booktabs}
\usepackage{amsmath}
\usepackage{amssymb}
\usepackage{graphicx}
\usepackage{xcolor}
\usepackage{listings}
\usepackage{multirow}
\usepackage{tikz}
\usetikzlibrary{positioning,arrows.meta}
\usepackage[hidelinks]{hyperref}
\usepackage{caption}
\lstdefinestyle{code}{
  basicstyle=\ttfamily\small,
  breaklines=true,
  frame=single,
  framerule=0.3pt,
  xleftmargin=1em,
  aboveskip=0.8em,
  belowskip=0.8em,
  columns=fullflexible,
  keepspaces=true
}

\newcommand{\tlaplus}{TLA\texorpdfstring{\textsuperscript{+}}{+}}

\title{\textbf{Can Code Specify a System Precisely Enough to Formally Verify It?}}

\author{
  Jean-Jacques Dubray\\
  \texttt{jdubray@gmail.com}
}
\date{July 4, 2026}

\begin{document}
\maketitle

\begin{abstract}
Formal verification is seldom applied to everyday production software,
because the cost of writing and maintaining a model has historically
exceeded the benefit. A companion study~\cite{companion} developed a
lower-cost alternative and evaluated it on a benchmark: specifications are
graded against traces captured from the running system. It found that when
large language models write the specifications, their reliability is
governed by the structure of the specification contract rather than by the
specification language. This paper evaluates both the alternative and that
finding on production software: the payment workflow of an
operational restaurant point-of-sale system, which must keep the register,
the payment terminal, and the payment processor in agreement. We report
three results. First, the core protocol is correct relative to a
hand-built, line-cited model under a failure model that we state precisely
(\S\ref{sec:whatchecked}). The audit identified seven failure-handling
gaps, nearly all with a common root cause; three were reproduced as real
executions,
and a patch closing them was re-checked with all failure gates enabled,
after which a follow-up patch closed a further defect that the re-check
itself exposed. Systematic extensions of the failure model (crash--restart,
stale reads, and two attempts) each identified the windows they were
designed to probe. Second, a single probe of the production payment sandbox
exposed a response-shape divergence that renders an entire recovery ladder
unreachable against the live API. The emulator-based audit was structurally
unable to detect this divergence, because the code and the emulator share
the same misreading; we characterize this as a correlated-oracle failure.
Third, the companion study's central finding replicates across seven models
from two vendors: contract structure, not language, governs what LLMs
specify reliably. The replication concerns the ordering of contracts and
the failure taxonomy rather than the absolute level: only the strongest
models reached the corpus ceiling, and the harder task restores
discriminating power that the benchmark had lost.
\end{abstract}

\section{Introduction}
\label{sec:intro}

The companion study~\cite{companion} extended
SysMoBench~\cite{sysmobench,sigops2026} with its first non-formal
specification backend (JavaScript in the SAM pattern, contributed upstream in
pull request~\cite{jssampr} and documented in a from-scratch
walkthrough~\cite{jssamwalkthrough}) and arrived, across four experiments and successive
adversarial audits, at the following ranking: on the tasks tested, LLM
specification quality was governed first by contract structure, second by
the prompt, and not at all by the specification language. This ranking
carried an explicit caveat. The tasks were a kernel spinlock with three
observable states and a small lock service, and the result was conditional
on replication against a system whose transition relation cannot be
inferred from its action names.

This paper provides that replication, and extends it. The target is not a
benchmark task but production code: the payment state-alignment workflow of
a live point-of-sale system, in daily use with real cards and real
transactions. The change of setting introduces a question that a benchmark
cannot pose. The question is not ``can models specify this system?'' but
``is this system correct?'', with consequences (lost charges, double
charges, and short-charged orders) measured in currency rather than in
benchmark points.

This paper is self-contained. The three specification
contracts under comparison are reproduced in Appendix~\ref{app:contracts},
every scoring definition is stated in \S\ref{sec:method}, and
\S\ref{sec:whatchecked} states exactly what the model checker verified,
under which failure modes, with which instantiation.

\begin{figure}[t]
\centering
\begin{tikzpicture}[
  agent/.style={draw, rounded corners, align=center, minimum height=1.05cm,
    inner sep=6pt, font=\small},
  human/.style={align=center, font=\small\itshape},
  ch/.style={-{Stealth}, thick},
  chd/.style={{Stealth}-{Stealth}, thick},
  lbl/.style={font=\scriptsize, align=center}]
\node[agent] (pos) {POS workflow\\(SAM state machine, \texttt{txState})};
\node[agent, right=4.8cm of pos] (finix) {Finix backend\\(transfer:
  \texttt{PENDING} $\to$\\\texttt{SUCCEEDED}/\texttt{FAILED}/\texttt{CANCELED})};
\node[agent, below=1.6cm of finix] (term) {PAX A920 terminal\\(Finix App)};
\node[human, below=1.6cm of pos] (staff) {staff:\\initiate / cancel / retry};
\node[human, left=1.0cm of term] (cust) {customer:\\tap / decline / cancel};
\draw[ch] ([yshift=11pt]pos.east) --
  node[lbl, above]{\texttt{POST /transfers} (idempotency key)\\
    \texttt{PUT /devices \{CANCEL\}}}
  ([yshift=11pt]finix.west);
\draw[ch] ([yshift=-11pt]finix.west) --
  node[lbl, below]{\texttt{GET /transfers/:id} --- poll, 2\,s (no webhooks)}
  ([yshift=-11pt]pos.east);
\draw[chd] (finix) -- node[lbl, right]{prompts / results} (term);
\draw[ch] (staff) -- (pos);
\draw[ch] (cust) -- (term);
\end{tikzpicture}
\caption{The system under study: three agents with no shared memory. Note
the absent edge: the POS and the terminal have \emph{no direct channel}.
The terminal's behaviour is visible to the POS only through the polled Finix
transfer state, which is what makes state alignment difficult.}
\label{fig:agents}
\end{figure}
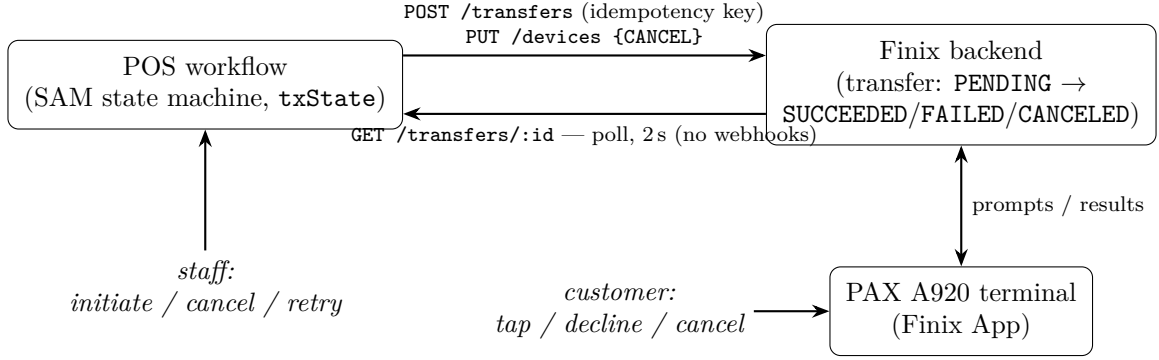

\paragraph{The system.}
The workflow keeps three agents in agreement; they share no memory and
communicate only through narrow, failure-prone channels
(Figure~\ref{fig:agents}): the POS (a SAM-pattern state
machine~\cite{sam,dubray2016infoq}), the PAX A920 Pro payment terminal
running the Finix App, and the Finix backend, which is polled every 2
seconds because no webhooks are available. The terminal is visible to the
POS \emph{only} through the polled Finix transfer state; no direct
POS--terminal channel exists. Four user-facing outcomes must be decided
correctly under every interleaving of customer taps, card declines, staff
cancels, device cancels, network failures, duplicate-key races, and a
180-second timeout: \textbf{Succeeded}; \textbf{Declined} (terminal, with
no automatic retry; the production-observed ``decline, then a subsequent
retry succeeds'' pattern is safe by design, because every attempt carries a
fresh idempotency key); \textbf{Error} (unreadable card, timeout, or
infrastructure failure); and \textbf{Canceled} (the customer pays cash
instead). The mapping is deliberately non-trivial: one machine state
(\texttt{DECLINED}) maps to three user outcomes through decline codes, and
Canceled is reachable through two machine states.

\paragraph{Provenance.}
The system under study is the point-of-sale system that the author built to
operate his restaurant, Hanuman Thai Cafe in Kirkland, Washington. It is
open source~\cite{kizo} and has a user manual~\cite{kizomanual}. The
payment workflow audited here is production code written by Claude models in
SAM semantics, and the author is also the author of the SAM pattern. The
auditing agents are Claude models, as are the verification agents that
built the reference models and the fix patch. The resulting circularity has
four components; the fourth is training-data contamination, which
\S\ref{sec:limitations} rules out by release dates for six of the seven
evaluated checkpoints and empirically, through a zero-hit beyond-prompt
artifact scan, for the seventh. \S\ref{sec:limitations} treats all four
components as first-class limitations. The countervailing evidence is the
same as in the companion study: the pipeline is automated, the trace corpus
is captured from real executions of the unmodified code, every claim is
accompanied by raw data, and the most consequential findings are
unfavorable to the author's own artifacts, namely a production system with
confirmed payment-handling gaps, the SAM contract again as the
worst-performing arm, and the author's own emulator was found to disagree with
the production API.

\paragraph{Contributions.}
\begin{enumerate}
\item \textbf{A scoped correctness verdict on a production payment
  protocol} (\S\ref{sec:correctness}), with the verification envelope
  stated precisely (\S\ref{sec:whatchecked}): seven failure-handling gaps,
  nearly all sharing one root cause, of which three were reproduced as real
  executions and five were isolated by single-failure-mode TLC
  counterexamples with line-cited traces; systematic failure-model
  extensions (crash--restart, stale reads, and cross-attempt retries) that
  identified the additional windows they were designed to probe; and a
  production-sandbox experiment that reclassified one gap and exposed a
  divergence between the code and reality that the emulator could not
  detect (\S\ref{sec:sandbox}).
\item \textbf{A verified fix} (\S\ref{sec:patch}): a 24-hunk patch based on
  a single principle, that no path abandons a possibly-live transfer
  without recording a recovery breadcrumb. The patch corrects every
  emulator reproduction and was then re-checked: every original
  counterexample is eliminated, and a strengthened invariant set holds with
  all failure gates enabled simultaneously. The re-check revealed one narrow
  bookkeeping defect in the patch itself; a validated follow-up (patch v2)
  closes it, together with the two remaining in-scope residuals.
\item \textbf{The methodology, transferred} (\S\ref{sec:method}): the
  companion study's apparatus (a pinned observable contract, mechanical
  replay in both languages, reference and mutation controls, and
  pre-committed interpretations) is carried to a system with 9 states, 10
  actions, and acceptor-level rewrites, at the cost of one new instrument
  (an instrumentation patch that captures the SAM dispatch as the transition
  boundary) and one restated inherited oracle commitment.
\item \textbf{A complete replication} (\S\ref{sec:comparison},
  \S\ref{sec:crossstudy}): the full 7-model $\times$ 3-arm $\times$
  $N{=}5$ grid across two vendors, in which all three of the companion
  study's mechanisms reappear, together with two new findings. The first is
  \emph{source-shape aggravation} of the SAM ceremony penalty: the weakest
  model imitates the production source's library idiom rather than the
  prompt's contract in five of five generations, producing plausible
  specifications that never execute. The second is the reappearance of a
  capability gradient within the bare arm on a task difficult enough to
  discriminate among models.
\end{enumerate}

\section{The system under study}
\label{sec:system}

The extraction (five files copied verbatim, together with a line-cited
protocol map in the artifact repository) scopes the study to the
single-terminal, single-payment lifecycle. The core is a 2{,}725-line SAM
module that implements a deterministic FSM over nine control states with
transition enforcement, plus the mechanisms that make the payment
path difficult to implement correctly:

\begin{description}
\item[Acceptor-level rewrites.] Two pre-FSM rewrites carry the protocol's
  most subtle semantics. A \texttt{TAP\_DECLINED} dispatched while in
  \texttt{CANCELLING} becomes an internal \texttt{CANCEL\_DECLINED} and is
  routed to \texttt{CANCELLED}, so that one action name reaches two targets
  without violating FSM determinism. A partial-approval guard rewrites a
  \texttt{TAP\_APPROVED} with \texttt{approvedAmount} $<$
  \texttt{amountCents} into a decline with code \texttt{PARTIAL\_PAYMENT};
  an overpayment is treated as a tip and accepted.
\item[Glitch tolerance.] Every model mutation is guarded by the expected
  post-transition state, so that FSM-rejected actions (for example, a late
  \texttt{TAP\_APPROVED} arriving in \texttt{COMPLETED}) are observable
  no-ops. This is the code's own anti-glitch invariant.
\item[The create-sale recovery ladder.] Transfer creation retries once on
  the \emph{same} idempotency key (Finix idempotency closes the ambiguity
  window), handles the 422 duplicate-key path by fetching the existing
  transfer, falls back to a last-resort device-cancel probe, and, if none
  of these resolves the transfer, writes a \texttt{pending\_terminal\_sales}
  breadcrumb and remains in \texttt{AWAITING\_VERIFICATION} for a later
  orphan sweep to resolve.
\item[The cancel ladder.] A device cancel may be preempted by a customer
  tap; the cancel path therefore re-probes Finix and must honour the charge
  when the tap wins (\texttt{CANCELLING} exits via \texttt{RECORDING}). The
  2-second poll continues to run during \texttt{CANCELLING} and can resolve
  the race first. The 180-second timeout is routed into the same cancel
  ladder.
\end{description}

The implicit correctness contract, extracted from the code's own comments
and mechanisms, forms the property set: \textbf{I1}, no lost charge (a
\texttt{SUCCEEDED} transfer is recorded, or a recovery mechanism can still
reach it); \textbf{I2}, no double charge; \textbf{I3}, cancel honoured
(\texttt{CANCELLED} $\Rightarrow$ the transfer did not succeed, or was
recorded instead); \textbf{I4}, declines are terminal (no automatic retry);
\textbf{I5}, no glitch writes; and \textbf{I6}, timeout liveness.
\S\ref{sec:whatchecked} states which of these were checked as invariants,
which through safety surrogates, and which were not model-checked at all;
the distinction is important, and an earlier draft of this paper did not
state it clearly.

\section{Methodology}
\label{sec:method}

The apparatus is inherited from the companion study~\cite{companion}; this
section restates everything on which this paper's claims depend and
describes the differences.

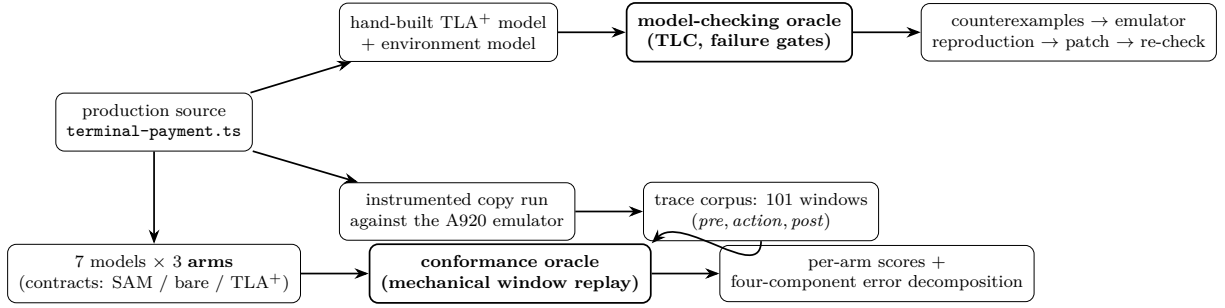
\begin{figure}[t]
\centering
\resizebox{\textwidth}{!}{%
\begin{tikzpicture}[
  box/.style={draw, rounded corners, align=center, font=\scriptsize,
    inner sep=5pt},
  oracle/.style={draw, thick, rounded corners, align=center,
    font=\scriptsize\bfseries, inner sep=5pt},
  arr/.style={-{Stealth}, thick}]
\node[box] (src) {production source\\\texttt{terminal-payment.ts}};
\node[box, above right=0.5cm and 1.4cm of src] (model)
  {hand-built \tlaplus{} model\\+ environment model};
\node[oracle, right=1.1cm of model] (tlc)
  {model-checking oracle\\(TLC, failure gates)};
\node[box, right=1.1cm of tlc] (cex)
  {counterexamples $\to$ emulator\\reproduction $\to$ patch $\to$ re-check};
\node[box, below right=0.5cm and 1.4cm of src] (harness)
  {instrumented copy run\\against the A920 emulator};
\node[box, right=1.1cm of harness] (corpus)
  {trace corpus: 101 windows\\$(\mathit{pre}, \mathit{action}, \mathit{post})$};
\node[box, below=1.5cm of src] (gen)
  {7 models $\times$ 3 \textbf{arms}\\(contracts: SAM / bare / \tlaplus{})};
\node[oracle, right=1.1cm of gen] (replay)
  {conformance oracle\\(mechanical window replay)};
\node[box, right=1.1cm of replay] (scores)
  {per-arm scores +\\four-component error decomposition};
\draw[arr] (src) -- (model);
\draw[arr] (model) -- (tlc);
\draw[arr] (tlc) -- (cex);
\draw[arr] (src) -- (harness);
\draw[arr] (harness) -- (corpus);
\draw[arr] (src.south) -- (gen.north);
\draw[arr] (gen) -- (replay);
\draw[arr] (replay) -- (scores);
\draw[arr] (corpus.south) to[out=-90, in=45] (replay.north east);
\end{tikzpicture}}
\caption{Study pipeline. The correctness deliverable (top lane) and the
language comparison (bottom lane) share the production source and the
emulator-captured trace corpus. The two \emph{oracles} are the boxed bold
nodes; the three \emph{arms} are the three specification contracts under
which each model generates, with everything else held fixed.}
\label{fig:pipeline}
\end{figure}

\paragraph{Terminology.}
An \emph{oracle} is the procedure that decides, for a given test item, what
counts as correct. This study uses two. The \emph{model-checking oracle} is
TLC exhaustively checking the invariants of \S\ref{sec:whatchecked} over the
hand-built model; its verdicts are counterexamples. The \emph{conformance
oracle} is mechanical replay of a captured window: the specification is
pinned to $\mathit{pre}$, $\mathit{action}(\mathit{data})$ is applied, and
the result is required to equal $\mathit{post}$ exactly; its verdicts are
per-window pass or fail. Oracles are instruments rather than ground truth:
they embed commitments and can themselves be incorrect. Two of this paper's
findings concern its own oracles, namely the correlated emulator of
\S\ref{sec:sandbox} and the tooling component of \S\ref{sec:comparison}. An
\emph{arm} is one condition of the controlled comparison, in the sense used
for a clinical trial: the set of specifications generated under one
specification contract (SAM, bare \texttt{next()}, or constrained
\tlaplus{}; Appendix~\ref{app:contracts}) with the model, prompt content,
corpus, and oracle held identical, so that a between-arm difference isolates
the effect of the contract. Figure~\ref{fig:pipeline} places both terms
within the study's architecture.

\paragraph{Observable contract and step semantics.}
The pinned observable state is the seven-field projection on which the four
outcomes depend: \texttt{txState}, \texttt{orderId}, \texttt{amountCents},
\texttt{transferId}, \texttt{declineCode}, \texttt{approvedAmountCents},
and \texttt{paymentId}. The step is the synchronous SAM dispatch: a window
is $(\mathit{pre}, \mathit{action}(\mathit{data}), \mathit{post})$, captured
at the entry of the acceptor chain and again after the full chain. The
asynchronous ladders constitute the \emph{environment}: they determine which
actions occur, and the scenario corpus exercises them. The specification
under test models the dispatch-level transition relation, which is precisely
what the acceptors together with the FSM implement. Because the code is
SAM-shaped, this boundary required no interpretation: the dispatch \emph{is}
the step, so the specification-versus-implementation gap that trace
conformance ordinarily polices is closed structurally \emph{for the model
core}. It is not closed for the environment model; \S\ref{sec:whatchecked}
states how that model is validated and what residual trust remains. The
oracle commitments identified in the companion study carry over unchanged
and unresolved: replay requires total, successor-producing actions (no-op
windows are treated as first-class), which penalizes enabling-condition
\tlaplus{} by construction, and its functional (branching-factor-one) form
is licensed only by the corpus's per-scenario determinism.

\paragraph{Trace corpus: real executions of the unmodified workflow.}
An instrumentation patch, applied to a copy so that the production source is
never modified, prepends a snapshot acceptor and appends an emitting
reactor. The harness drives the actual \texttt{createTerminalPaymentWorkflow}
against the repository's own PAX A920 emulator, which intercepts the Finix
HTTP surface in-process and can script declines, 422 races, cancel races,
and partial approvals, using a disposable database and a reduced timeout for
the timeout scenario. Corpus v1 contains 75 windows from 17 scenarios.
Corpus v2 adds 6 scenarios (26 windows, for a total of 101), so that each
acceptor-rewrite rule is detected by three windows rather than one; two of
the new scenarios are themselves emulator-captured evidence of defects
(\S\ref{sec:gaps}). One consequence should be stated plainly: every window
is \emph{emulator}-generated, so the language comparison measures fidelity
to the code as emulated. This is the appropriate target for a controlled
between-arm comparison, since all arms share it, but \S\ref{sec:sandbox}
shows that the emulator and reality diverge in at least one place, so a
per-window ``pass'' must not be read as ``matches Finix.''

\paragraph{Reference models and controls.}
Correctness is judged by a hand-built, line-cited \tlaplus{} model of the
dispatch relation (every operator total, with the two rewrites folded in)
composed with an explicit environment model comprising the Finix transfer
lifecycle, the create ladder with its attempt phases, the poll loop gated
exactly as the code gates it, the cancel probe, the customer and staff
actions, and the orphan sweep, which is enabled if and only if a breadcrumb
row exists; this last condition is the detail on which the defects turn.
\S\ref{sec:whatchecked} gives the instantiation and the full configuration
table. The comparison arms are validated by the control battery:
hand-written reference specifications in all three languages score 101/101,
and two rule-drop mutations (dropping the cancel rewrite, and dropping the
partial guard) each fail exactly their three target windows, identically in
all three languages. The replay therefore discriminates symmetrically, at
rule granularity.

\paragraph{Comparison design and scoring, in full.}
All three arms are in \emph{derivation} mode, with no semantics
block: the model receives the module shape, the seven observable fields, the
action alphabet with data schemas, a description of the replay mechanism,
and the 3{,}000-line source packet, and must derive the transition relation
itself. The arms are the deployed SAM contract, the bare \texttt{next(state,
action, model)} contract, and the constrained \tlaplus{} contract
(Appendix~\ref{app:contracts}). There are $N{=}5$ generations per model per
arm across four Claude models (Haiku 4.5, Sonnet 4.6, Opus 4.8, and Fable 5)
and, in a cross-vendor panel (\S\ref{sec:crossvendor}), three Mistral
models. Each window scores \emph{pass}, \emph{fail}, or \emph{unscoreable},
where unscoreable denotes a window that could not be driven to a verdict
because of a module load failure, a dead dispatch, or a TLC deadlock on a
no-op window. Two quantities are computed per generation, both fixed in the
study plan before the grid was run: the \emph{unconditional} pass rate
(passes over all 101 windows) and the \emph{conditional} pass rate (passes
over scoreable windows). The analysis of record is a single uniform re-score
of all 105 generations on corpus v2, run identically for every model and
arm; no per-model adjudication was performed (\S\ref{sec:decomposition}
reports the one tooling artifact that this policy exposed). The run paused
at 15 of 60 Claude cells because of an API budget limit and was subsequently
resumed to completion; all analyses are over the full grid.

\section{Correctness of the implementation}
\label{sec:correctness}

\subsection{Scope of the verification}
\label{sec:whatchecked}

This subsection makes the word ``verified'' in this paper checkable. It
states the model, the instantiation, the failure model, the properties, and
the exact TLC runs. Everything described here is included in the artifacts
(\texttt{reference/}, seven pre-fix and seven core post-fix configurations
--- plus two additional post-fix diagnostic runs --- and annotated
counterexamples).

\paragraph{Model and instantiation.}
\texttt{FinixPOS.tla} (157 lines) models the dispatch relation: one total
operator per SAM action, with FSM-rejected actions as explicit no-ops and
the two acceptor rewrites folded in; every operator cites the source lines
it models. \texttt{FinixPOSEnv.tla} (373 lines) composes this with the
environment: the Finix transfer lifecycle, the create ladder (attempt
phases 0--4, with a \texttt{createdBy} marker distinguishing ``my
response'' from ``422 duplicate''), the 422 handler, the poll loop gated as
\texttt{pollTick} gates it, the cancel probe, the record step, the customer
and staff actions, and the orphan sweep, enabled if and only if a breadcrumb
exists. The instantiation uses one order, one idempotency key, one transfer
id, one payment id, and two amounts (\texttt{REQ\_AMT} $=$ 10 and
\texttt{PARTIAL\_AMT} $=$ 5). This is not an arbitrary reduction: the
observable contract compares identifiers only for equality and presence, and
amounts only through the single comparison \texttt{approvedAmount} $<$
\texttt{amountCents}, so one identifier per role and two amounts exhaust the
distinctions the protocol can make \emph{within a single attempt}.
Cross-attempt behaviour is precisely what a single-attempt model cannot
observe, which is why \S\ref{sec:extmodel} adds a two-attempt model rather
than assuming the abstraction to be sound; that model identified gap 10. The
specification is $\mathit{Init} \wedge \Box[\mathit{Next}]_{\mathit{vars}}$
with \emph{no fairness assumptions}. TLC runs with deadlock checking
disabled, because a completed single-attempt run is a terminal state by
design. Every run explores the complete reachable graph (43--103 distinct
states per pre-fix configuration, in under a second). The protocol is small
once projected onto the observable contract, which is itself a finding about
where its difficulty resides, and it means that exhaustiveness is
inexpensive here rather than notable.

\paragraph{The failure model.}
The \emph{base} environment already contains the failure modes that the code
was designed to handle: create requests and responses can be lost (a request
that is lost from the POS's perspective may still be processed by Finix
afterwards, which is the idempotency ambiguity), the same-key retry, the 422
duplicate-key path, the last-resort probe, every customer and staff
interleaving, and the tap-beats-cancel race. Five \emph{gates}
(\textsc{constants}) add the failure modes that the code's periphery must
survive: \texttt{EARLY\_CANCEL} (a staff cancel from \texttt{INITIATING}),
\texttt{FETCH\_CAN\_FAIL} (the 422-path status fetch fails),
\texttt{PARTIAL\_POSSIBLE} (Finix approves less than requested),
\texttt{DB\_CAN\_FAIL} (the breadcrumb \textsc{insert} fails), and
\texttt{CANCEL\_BLIND\_POSSIBLE} (the cancel response and the probe both
miss a \texttt{SUCCEEDED} transfer). Pre-fix, gates are enabled \emph{one
per configuration}, so that each counterexample isolates one mechanism; a
pre-fix run with all gates enabled would add nothing, since each gate
already violates an invariant individually. Gate \emph{interactions} are
therefore explored only post-fix, where the patched model holds with all
five gates enabled simultaneously (Table~\ref{tab:tlcruns}).
\S\ref{sec:extmodel} extends the model along three axes that a per-channel
enumeration showed to be missing: process crash--restart, stale reads, and
cross-attempt composition.

\paragraph{The checked properties.}
All checked properties are \emph{safety} invariants over the composed state;
Figure~\ref{fig:invariants} gives the operative formulas verbatim. Against
the informal contract I1--I6 of \S\ref{sec:system}:

\begin{description}
\item[I1 (no lost charge)] is checked as a safety surrogate: no reachable
  state has a \texttt{SUCCEEDED} transfer, a terminal
  \texttt{DECLINED} or \texttt{CANCELLED} POS, no payments row, and no
  breadcrumb. The breadcrumb serves as a \emph{recoverability witness}. The
  corresponding liveness property, that the sweep eventually resolves every
  breadcrumb, is \emph{not} checked; under no fairness assumptions it could
  not be. What is checked is that recovery is never rendered impossible.
\item[I2 (no double charge)] cannot be probed by a single-attempt model
  beyond bookkeeping consistency (\texttt{Inv\_NoPartialRecorded}); within
  one attempt it is enforced structurally by the fixed idempotency key and
  the \textsc{on conflict} guard. The two-attempt model of
  \S\ref{sec:extmodel} checks it directly, as three invariants: at most one
  \texttt{SUCCEEDED} charge, at most one payments row, and detection of any
  double charge.
\item[I3 (cancel honoured)] is checked directly.
\item[I4 (declines terminal)] is not model-checked, because retry gating is
  staff-mediated environment behaviour rather than dispatch semantics. It is
  exercised by the trace scenarios (a decline followed by a fresh-key retry)
  and encoded in the two-attempt model's retry gating.
\item[I5 (no glitch writes)] holds by construction in the total dispatch
  model, since rejected actions are no-ops, and is exercised empirically by
  the corpus's eight observed no-op combinations, which leave the state
  bit-identical.
\item[I6 (timeout liveness)] is not checked. The model is untimed, and the
  180-second timeout is merged with the staff cancel, because the code
  dispatches the same \texttt{CANCEL\_PAYMENT}. The one liveness-related
  defect that the study found, the gap-4 deadlock (a state with no outgoing
  transition), was detected by expressing reachability of that state as a
  safety property (\texttt{Inv\_VerificationHasBreadcrumb}).
\end{description}

A type-correctness invariant (\texttt{TypeOK}) is checked in every
configuration alongside the property invariants.

\begin{figure}[t]
\begin{lstlisting}[style=code]
Inv_I1_NoLostCharge ==
  ~( /\ finixState = "SUCCEEDED"
     /\ txState \in {"DECLINED", "CANCELLED"}
     /\ ~paymentRecorded
     /\ ~breadcrumb )

Inv_I3_CancelHonoured ==
  (txState = "CANCELLED") => (finixState # "SUCCEEDED" \/ paymentRecorded)

Inv_VerificationHasBreadcrumb ==   \* the gap-4 deadlock, phrased as safety
  (txState = "AWAITING_VERIFICATION") => breadcrumb

Inv_NoPartialRecorded ==           \* I2-adjacent bookkeeping
  paymentRecorded =>
    (approvedAmountCents = NUMNONE \/ approvedAmountCents >= REQ_AMT)

Inv_RecordingHasTransferId        \* gap-7 invariant (post-fix set only;
                                  \* fails pre-fix, shown live in a v2 scenario)
\end{lstlisting}
\caption{The checked invariants, verbatim from \texttt{FinixPOSEnv.tla} /
\texttt{FinixPOSFixedEnv.tla}. All are safety properties;
\S\ref{sec:whatchecked} states what was \emph{not} checked (liveness) and
why.}
\label{fig:invariants}
\end{figure}

\begin{table}[t]
\centering
\caption{Every TLC run backing the correctness claims. All runs are
exhaustive over the complete reachable graph; ``gates'' names the failure
modes enabled beyond the base environment (which always includes lost/late
create responses, the same-key retry, the 422 path, and all customer/staff
interleavings). Counterexample depths are 6--9 steps; annotated traces ship
in \texttt{counterexamples/}.}
\label{tab:tlcruns}
\small
\begin{tabular}{llllr}
\toprule
Configuration & Gates enabled & Invariants & Result & States \\
\midrule
\multicolumn{5}{l}{\emph{Pre-fix} (\texttt{FinixPOS} + \texttt{FinixPOSEnv})} \\
\texttt{clean} & none & TypeOK, I1, I3, deadlock, partial & \textbf{holds} & 103 \\
\texttt{bug1\_earlycancel} & \texttt{EARLY\_CANCEL} & I1 & violated (7 steps) & 43--103 \\
\texttt{bug2\_fetchfail} & \texttt{FETCH\_CAN\_FAIL} & I1 & violated (7) & \\
\texttt{bug3\_partial} & \texttt{PARTIAL\_POSSIBLE} & I1 & violated (7) & \\
\texttt{bug3b\_partialrec} & \texttt{PARTIAL\_POSSIBLE} & partial & violated (9) & \\
\texttt{bug4\_dbfail} & \texttt{DB\_CAN\_FAIL} & deadlock & violated (6) & \\
\texttt{bug5\_blindcancel} & \texttt{CANCEL\_BLIND} & I3 & violated (8) & \\
\midrule
\multicolumn{5}{l}{\emph{Post-fix} (\texttt{FinixPOSFixed} + \texttt{FinixPOSFixedEnv})} \\
\texttt{fixed\_clean} & none & all five (incl.\ gap-7) & \textbf{holds} & 131 \\
\texttt{fixed\_bug1..bug5} & one each & per-gap set & \textbf{holds} & 135--239 \\
\texttt{fixed\_allfail} & \textbf{all five} & I1, I3, deadlock, gap-7 & \textbf{holds} & 411 \\
\texttt{fixed\_allfail\_partialbook} & all five & partial & violated (6) $\to$ patch v2 & \\
\texttt{fixed\_bug4\_strict} & \texttt{DB\_CAN\_FAIL} & I1 (no db-exemption) & violated (8), inherent & \\
\midrule
\multicolumn{5}{l}{\emph{Patch v2} (\texttt{FinixPOSFixedV2} + env)} \\
\texttt{fixedv2\_allfail*} & all five & full set incl.\ partial & \textbf{holds} & 376 \\
\midrule
\multicolumn{5}{l}{\emph{Extensions} (\S\ref{sec:extmodel}; own modules, pre/post-fix by a \texttt{PATCHED} constant)} \\
\texttt{ext\_crash\_*} & crash--restart & I1 & violated pre \& post (cases 8/9/9b) & 24--52 \\
\texttt{ext\_stale\_*} & 1-step stale reads & I1 & violated pre, \textbf{holds} post & 17--18 \\
\texttt{ext\_twoatt\_*} & two attempts & I2 (money/booked/visible) & Table in \S\ref{sec:extmodel} & 21--38 \\
\bottomrule
\end{tabular}
\end{table}

\paragraph{Validation of the environment model, and its limits.}
The dispatch core is validated mechanically: the trace corpus replays
101/101 against the hand-written reference, and the observed
$(\mathit{pre.txState}, \mathit{action})$ combinations behave exactly as the
extraction predicts, no-ops included. The environment model has no
equivalent mechanical check, because the corpus captures dispatch windows
rather than ladder internals, yet the correctness verdict depends on it. Its
validation is threefold and weaker: every environment action cites the
source lines it models; three of the predicted gaps were reproduced as real
executions of the unmodified code, confirming the environment model's
predictions in detail, including the predicted NULL transfer id of gap 7;
and the post-fix corpus discriminates exactly the six windows whose
semantics the patch changed, and no others. This does not rule out an
environment behaviour on which the model, the code, and the emulator agree
but which reality contradicts, which is precisely the class of defect that
\S\ref{sec:sandbox} found once, by testing against the production sandbox
rather than the correlated triple. One known coverage gap is recorded: the
patch's new \texttt{AWAITING\_VERIFICATION} staff-cancel exit appears in the
TLC model but not in the trace corpus.

\paragraph{The verdict.}
With the base failure model and no gates enabled, TLC exhaustively verifies
the four pre-fix invariants over the complete state graph, including the
difficult cases: tap-beats-cancel resolved in favour of the charge, the full
422 recovery ladder, sweep resolution of verification-pending sales, and
late or duplicate-action no-ops. All four user outcomes are decided
correctly on their main paths. Each of the five gates, enabled alone,
violates an invariant; these correspond to five of the gaps of
\S\ref{sec:gaps}. Process
crash--restart and terminal power loss are treated as accepted operational
risk, which is a deployment decision rather than an oversight: this is a
staff-present restaurant POS with small tickets, cash remediation is always
available, and the \texttt{payment\_events} log preserves the idempotency
key of every attempt for manual reconciliation. The crash boundary was
examined rather than omitted: \S\ref{sec:extmodel} reports the
crash--restart model and the three cases it identified before they were
classified.

\subsection{Identified defects: seven gaps, nearly all with a common root cause}
\label{sec:gaps}

Every violation occurs in the failure-handling periphery, and nearly all
reduce to the same design omission: \emph{a path abandons a possibly-live
transfer without leaving the idempotency-key breadcrumb (or scoped guard) on
which the recovery machinery depends.} The two exceptions (gaps 6 and 7)
concern transfer-id selection rather than the missing breadcrumb.

Table~\ref{tab:bugs} lists the seven originally identified gaps in eight
rows; gaps 3 and 3b are two manifestations of the single partial-approval
gap. The numbering continues past this table deliberately: cases 8, 9, and
9b are the crash--restart findings and gap 10 is the cross-attempt finding,
all in \S\ref{sec:extmodel}, so that every numbered finding in the artifacts
has exactly one name in the paper.

\begin{table}[ht]
\centering
\caption{The seven originally-identified gaps (eight rows; 3/3b are one
gap). ``Emu.'' = reproduced as a real execution of the unmodified code
against the repository's Finix emulator (not the production API; see
\S\ref{sec:sandbox}). Each TLC counterexample is isolated by a single
failure-mode gate, with line-cited traces in the artifacts. Gap 2 was
subsequently reclassified by the real-sandbox experiment.}
\label{tab:bugs}
\small
\begin{tabular}{clcc}
\toprule
\# & Gap & Found by & Emu. \\
\midrule
1 & Cancel during \texttt{INITIATING} strands a live, tappable transfer:
    charged at Finix, & TLC & \textbf{yes} \\
  & \texttt{CANCELLED} at POS, no breadcrumb --- invisible to every sweep
    & & \\
2 & 422 duplicate-key + failed status fetch declines with no breadcrumb
    row & TLC & \\
3 & Partial approval declined, but the settled partial charge is never
    voided & TLC & \\
3b & Partial guard covers only \texttt{AWAITING\_TAP}: partials via
    \texttt{CANCELLING} or sweep & TLC & \textbf{yes} \\
   & resolution are recorded --- order marked paid below total & & \\
4 & Breadcrumb INSERT failure still enters \texttt{AWAITING\_VERIFICATION},
    which then & TLC & \\
  & has no exit: workflow deadlocked, order locked & & \\
5 & ``Conservative'' cancel (cancel throws and probe fails over a
    \texttt{SUCCEEDED} & modeling & \\
  & transfer) confirms cancellation over a real charge; no sweep covers
    \texttt{CANCELLED} rows & & \\
6 & Cancel probe prefers the cancel response's (possibly bogus) transfer
    id over the & harness & \\
  & polled one & & \\
7 & \texttt{resolveVerification} drops the passed \texttt{transferId}:
    recovered payment row has & harness & \textbf{yes} \\
  & \texttt{finix\_transfer\_id = NULL}, disabling the double-record guard
    for that row & & \\
\bottomrule
\end{tabular}
\end{table}

Three of the gaps warrant a fuller description. \textbf{Gap 1}, the most serious, requires no
optional failure mode at all, only the FSM-permitted staff cancel from
\texttt{INITIATING} racing the in-flight \texttt{POST /transfers}. The
device cancel finds nothing to cancel, the transfer is created afterwards,
its \texttt{TRANSFER\_CREATED} is silently FSM-rejected in the cancel
states, and the customer can still tap the live transfer. It was reproduced
as a real execution of the unmodified code. \textbf{Gaps 3b and 7} were
confirmed by the corpus itself: the two adversarial v2 scenarios are real
executions in which the unmodified workflow recorded \$16.80 charges against
\$21.80 orders and marked the orders paid, with the sweep-resolution variant
additionally writing the payment row with a NULL transfer id, exactly as the
model checker predicted.

\paragraph{Environment assumptions, checked against the Finix documentation.}
Partial authorization (gaps 3 and 3b) is a real, merchant-configurable Finix
feature (\texttt{default\_partial\_authorization\_enabled}); the
counterexample class is production-plausible, and the merchant profile is
worth verifying. The ordering of the device cancel relative to the in-flight
create (the enabling assumption for gap 1) is undocumented; the model's
reading, that a cancel does not fence a still-propagating create, is the
conservative one, and the defensive fix is inexpensive in any case. One
model-versus-code caveat is recorded in the artifacts: response bodies are
modelled as current-state snapshots, and \S\ref{sec:extmodel} relaxes this
with an explicit stale-read extension.

\subsection{The production-sandbox experiment and the correlated oracle}
\label{sec:sandbox}

Everything above (every trace, every emulator reproduction, and the patch's
scenario validation) executes against the repository's own Finix emulator,
which was written in the same repository, by the same author-and-model
lineage, and embodies the same understanding of Finix as the code under
audit. If the code and the emulator share a misunderstanding, both agree and
the audit cannot detect it. This is a \emph{correlated oracle}, and it is
not hypothetical: a single probe of the production Finix sandbox found an
instance.

The experiment (with evidence and redacted transcripts in the artifacts)
proceeded as follows: against \texttt{finix.sandbox-payments-api.com}, and
pinning the same \texttt{Finix-Version} that the production adapter sends, a
transfer was created and the same \texttt{idempotency\_id} was replayed. The
production API does return 422 on a duplicate key, but the 422 body carries
the existing transfer id only in \texttt{\_links.transfer.href} and in the
human-readable message. There is no \texttt{errors[0].transfer} field, which
is the only shape that the production parser (and, faithfully, the emulator)
recognizes. Replaying the captured production body through the actual
production adapter raises a generic \texttt{Error} rather than the typed
duplicate-key error. Consequently the entire duplicate-key recovery ladder,
including gap 2's blind-decline catch and the fast-record-on-\texttt{SUCCEEDED}
branch, is \textbf{unreachable against the production API} for the pinned
version: a real duplicate 422 falls through the generic error path to the
Case-4 breadcrumb.

There are three consequences. First, \textbf{gap 2 is reclassified}: the
blind-decline lost-charge scenario is unreachable against the production
API, and its risk is absorbed by the breadcrumb path, leaving a liveness and
usability regression (a tap-succeeded duplicate remains in verification
instead of recording immediately) and a device-scoped last-resort cancel
that can abort an unrelated prompt. Second, \textbf{the methodological
finding is more important than the gap itself}: the code and its emulator
agreed on a response shape that reality does not produce. This is the
correlated-oracle problem, observed on the first test against the production
API, and it is the reason every reproduction in this paper is labelled as
\emph{emulator}-reproduced. Third, the defensive fix (parsing the id from
\texttt{\_links.transfer.href}) makes the now-breadcrumbed ladder reachable
and correct; it is included in patch v2 (\S\ref{sec:patch}), together with a
correction of the emulator's 422 body to the production shape. The caveats
should be stated precisely: the validation used a card-not-present transfer
(no activated A920 exists on this sandbox), the sandbox is not production,
and the response shapes may be version-dependent, so the claim is scoped to
the deployment's pinned version.

\subsection{Extensions to the failure model: crashes, stale reads, and retries}
\label{sec:extmodel}

The original five gates concerned interleavings and error-response shapes. A
systematic per-channel enumeration (every RPC $\times$ \{lost request, lost
response, duplicate, stale read, error response\}, together with a process
crash between any two steps; the full table is included in the artifacts)
identified three uncovered rows. Each was then modelled.

\paragraph{Crash--restart (accepted operational risk; boundary examined).}
A crash model with faithful rehydration semantics (dehydration on every
transition, restart fast-forwarding only \texttt{AWAITING\_TAP} rows, and
the stale-\texttt{INITIATING} cleanup) identifies three cases in which a
charge can survive the process's termination undetected: \textbf{case 8},
a crash during \texttt{INITIATING}, where the cleanup marks the row
\texttt{CANCELLED} without a probe and the row carries no transfer id;
\textbf{case 9}, a crash in \texttt{RECORDING} between Finix success and
the local record; and \textbf{case 9b}, a crash in \texttt{CANCELLING}.
The designed case, a crash in \texttt{AWAITING\_TAP}, recovers correctly
both pre- and post-fix. The root cause is the same abandon-without-breadcrumb
pattern as in the earlier gaps, reached through the crash path: rehydration loads
non-\texttt{AWAITING\_TAP} rows but cannot drive them, and no sweep covers
the resulting orphaned rows. In accordance with the deployment's risk
posture (\S\ref{sec:whatchecked}), these are documented as accepted
operational risk rather than defects; the counterexamples are retained, and
a single optional hardening (a sweep over crashed non-terminal rows keyed by
the always-persisted idempotency key) closes all three if the posture
changes.

\paragraph{Stale reads (in scope; pre-fix violation, post-fix verified).}
Allowing one-step-stale \texttt{GET /transfers} responses yields a lost
charge pre-fix through the cancel ladder alone, a strictly weaker
precondition than that of gap 5: staleness, rather than an error, is
sufficient. Post-fix, the \texttt{CANCELLED}-rows sweep provably recovers it
(exhaustively), and poll-loop and 422-path staleness are self-healing.

\paragraph{Two attempts (in scope; the anticipated composition occurs).}
A two-attempt model with the code's actual retry gating confirms the
double-charge composition mechanically pre-fix: a blind decline over a live
transfer, a customer tap, an unblocked staff retry, and a second success
produce two \texttt{SUCCEEDED} transfers for one order, the first invisible
to every sweep. The books never double-count, which is why the incident is
undetected. Post-fix, the no-double-charge property holds exhaustively on
the dashboard path (breadcrumb, 409 pending-row guard, sweep, and
paid-order gating). However, the model also identified \textbf{gap 10}: the
\emph{counter} entry path lacks the dashboard's 409 guard, so a post-fix
counter retry can still race the sweep into a second real charge. This is
detected by the sweep's covered-order guard, which emits a warning-level log
entry, with the money movement left to the manual refund path. This case
occurs even when all equipment is working normally, and is in scope; patch v2 lifts the
guard into the shared entry point. One assumption is recorded: the model
encodes ``a staff retry only on a visibly unpaid order''; removing this
assumption makes even the dashboard path violate no-double-charge, but a
staff member re-charging a visibly paid order is operator error and is out
of scope.

\subsection{Verifying the fix}
\label{sec:patch}

A 24-hunk patch (delivered as a reviewable diff together with an annotated
companion, and deliberately \emph{not} applied to the production source)
implements a single idea throughout: no path abandons a possibly-live
transfer without recording a breadcrumb. A shared helper closes gaps 1, 2,
and 4; the partial-approval guard is extended to \texttt{CANCELLING} and
\texttt{AWAITING\_VERIFICATION}, with partial rejections surfaced for manual
refund (gaps 3 and 3b); \texttt{AWAITING\_VERIFICATION} gains a fail-closed
decline and a staff-cancel exit (gap 4); a new sweep covers
\texttt{CANCELLED} rows that carry a transfer id (gap 5); the cancel probe
checks both candidate ids (gap 6); and the transfer id is threaded through
verification recovery (gap 7).

Refunds are deliberately absent from the patch, as an access-control
decision: the POS has no per-employee refund authorization, so a POS-issued
refund would execute with system authority, granting every employee refund
power, whereas the Finix dashboard keeps that authority behind its own login
and roles. The constraint is organizational, and the formal apparatus could
not have identified it; an earlier draft of the patch issued automatic
refunds until the owner supplied this constraint. This parallels the
correlated-oracle observation of \S\ref{sec:sandbox}. Removing the refund
call also eliminated an assumption: the earlier draft's double-refund
protection depended on untested Finix reversal idempotency (there is no
reversals endpoint in the emulator, and it was not exercised in the
sandbox), and TLC confirmed that the protection held \emph{if and only if}
that assumption did. The manual-refund policy removes the property entirely:
the surfaced payment-error record is now the recoverability witness.

\paragraph{Scenario validation.} The exact harness scenarios that produced
the emulator reproductions were re-run against the patched copy, and every
outcome is corrected: there is no stranded transfer and the breadcrumb is
present; the recovered payment row carries the real transfer id; and the
partial scenarios end in \texttt{CANCELLED} or \texttt{DECLINED} with
nothing recorded and the charge surfaced for manual refund, while a
recording proxy placed in front of the emulator observed no refund or
reversal requests. All scenario assertions pass, and the type check is clean
against the project configuration.

\paragraph{Re-checking the model.} Scenario testing is not verification, so
the patched semantics were modelled and the full configuration battery
re-run (Table~\ref{tab:tlcruns}): every original per-gap counterexample is
eliminated, and a strengthened invariant set, including the gap-7 invariant
that fails pre-fix, holds with all five failure gates enabled simultaneously
over the complete graph. The patch adds transitions (a staff-cancel exit, a
new sweep, and widened guards), so the re-check also searched for
regressions and found one: on the \emph{recovery} paths, the pre-existing
fast-record helper writes the payments row before the widened guard can
reject it, so a recovery-discovered partial approval is recorded, then
declined and surfaced for manual refund. The order is marked paid at the
partial amount while the charge is flagged for return, so the books are
incorrect in either case. Pre-fix, this same path was gap 3 itself, so the
patch improves it but does not fully resolve it. A second, inherent residual
is documented: if the local database is unavailable, the fail-closed decline
can still lose a late-materializing charge, because no database-based
recovery can exist when the database is the failed component (an equipment
failure, under the same accepted posture as in \S\ref{sec:extmodel}, with
the warning-level log as the manual path).

\paragraph{Patch v2, validated.} A cumulative follow-up closes the three
in-scope residuals: the gap-10 guard is lifted into the counter entry point;
the production-API 422 parser (\S\ref{sec:sandbox}) and the corrected
emulator body are added; and the partial check is moved before the database
write, with both recovery sweeps modified to reject partials. The last item
was added because the v2 re-check found that the v1 \texttt{CANCELLED}-rows
sweep would re-record a partial that the guard had rejected. TLC holds over
the full failure-enabled graph (376 distinct states, all gates enabled, and
\texttt{Inv\_NoPartialRecorded} included), and the reachability tests pass,
including a recorder assertion that no refund API call is ever made.

\paragraph{Post-fix corpus and generations.} Re-running all 23 study
scenarios against the patched instrumented workflow yields a 99-window
post-fix corpus, with 23 of 23 scenarios passing under database-level
assertions. The pre-fix reference specification, replayed on this corpus,
fails exactly the six windows whose semantics the patch changed, and no
others. Ten fresh bare-arm generations (Fable~5 and Opus~4.8, $N{=}5$ each)
against the patched source all score 99/99: the ceiling result transfers to
the modified system, and two models independently re-derive the patched
semantics from the source alone.

\section{The language comparison at production scale}
\label{sec:comparison}

\begin{table}[ht]
\centering
\caption{Full Claude grid: mean pass rate over $N{=}5$ generations per
cell, corpus v2 (101 windows), derivation mode. Each cell shows the
unconditional (conditional) pass rate, the two pre-committed quantities of
\S\ref{sec:method}. Every difference between the unconditional and
conditional values is due to guard-idiom unscoreable windows, never to
incorrect post-states.}
\label{tab:arms}
\small
\begin{tabular}{lccc}
\toprule
Model & SAM contract & bare \texttt{next()} & \tlaplus{} (constrained) \\
\midrule
Claude Fable 5    & \textbf{100 (100)} & \textbf{100 (100)} & \textbf{100 (100)} \\
Claude Opus 4.8   & 98.8 (98.8) & \textbf{100 (100)} & 98.8 (99.6) \\
Claude Sonnet 4.6 & 98.8 (98.8) & 99.4 (99.4) & 88.3 (98.3) \\
Claude Haiku 4.5  & \textbf{0.0} (0.0) & 95.8 (95.8) & 79.6 (97.1) \\
\bottomrule
\end{tabular}
\end{table}

\paragraph{Statistics.} Generation-level exact permutation tests, read
descriptively in accordance with the companion study's analysis of record,
give the following. Opus and Fable show no significant contrasts, since both
are at the ceiling, where effect ordering is unidentifiable. For Sonnet,
bare $>$ \tlaplus{} and SAM $>$ \tlaplus{}, both at the permutation floor
($p{=}.0079$). For Haiku, all three contrasts are at the floor, including
bare $>$ SAM at $\Delta{=}{+}.958$, the largest effect in either study.

\paragraph{The bare-\texttt{next()} ceiling replicates, and a capability
gradient reopens within it.} The bare contract is best or tied for every
model. However, unlike the benchmark tasks, which the bare contract
saturated for all four models, the production protocol discriminates
\emph{within} the bare arm: Fable 100\%, Opus 100\%, Sonnet 99.4\%, and
Haiku 95.8\%. Every bare miss is semantic, on the acceptor rules that lie
outside the FSM table. The clearest example is one Sonnet bare generation
that transcribed the FSM table verbatim and missed only the override
acceptor. This directly addresses the companion study's saturation concern:
discriminating power is restored by harder tasks rather than by heavier
contracts.

\paragraph{The \tlaplus{} deficit remains a replay-contract artifact across
the Claude grid.} Every remaining Claude \tlaplus{} loss is a guard-idiom
deadlock: unscoreable rather than incorrect. Conditional \tlaplus{} fidelity
is 97--100\% for every Claude model, Haiku included: its \tlaplus{} arm
reaches 97.1\%, even though its SAM (0.0\%) and bare (95.8\%) arms fall below
that band for unrelated reasons --- ceremony failure and semantic misses,
respectively. Nothing in this grid shows a model writing incorrect
\tlaplus{} transitions that it writes correctly in JavaScript, or the
converse. The oracle-paradigm commitment identified in
the companion study is essential here as well; the conditional score is
co-reported as the paradigm-fair quantity throughout, and
\S\ref{sec:crossvendor} shows where the claim does \emph{not} extend.

\paragraph{The ceremony penalty returns in a stronger form: source-shape
aggravation, five of five.} All five Haiku SAM generations share an
identical dead-wiring signature: they copied the production source's
\texttt{['NAME', fn]} tuple idiom for wiring actions, whereas the library
contract expects plain functions. As a result, intents never fire and every
dispatch is dead, giving 0/101 despite mostly correct transition logic: the
specifications reproduce the cancel rewrite and the partial guard, the two
hardest rules. Five of five rules out sampling noise: this is systematic
imitation of the source in preference to the instruction, and it means that
a SAM-shaped source \emph{aggravates} the SAM contract's penalty rather than
mitigating it. The richer the contract surface, the more ways a
plausible-looking specification can fail to execute at all. The imperfect
Opus and Sonnet SAM generations make the complementary point: their failing
windows are exactly the two rewrite rules that the mutation controls target,
that is, the hardest semantic content rather than ceremony.

\paragraph{A four-component error decomposition, and the scoring policy
behind it.}
\label{sec:decomposition}
Every lost window in the grid is attributable to exactly one of four
components. The \emph{semantic} component consists of misses of the two
acceptor rules and is language-independent (Haiku's bare and \tlaplus{}
specifications miss the same partial-guard windows, three to four across
generations). The
\emph{replay-contract} component consists of all remaining Claude \tlaplus{}
losses, which are guard-idiom deadlocks with conditional fidelity intact.
The \emph{ceremony} component consists of the SAM wiring failures, up to
total failure. The \emph{tooling} component is as follows: in the superseded
v1 scoring pass, two Fable \tlaplus{} windows were reported unscoreable;
they pass in 2.4 seconds on an isolated re-run and pass in the uniform v2
rescore of all 105 generations. These were transient TLC/JVM failures under
machine load. To address an adjudication concern directly: no cell was
individually re-adjudicated. The analysis of record is the single uniform v2
rescore applied identically to every generation of every model, under which
Fable's grid is perfect across all three arms, the only model to achieve
that. The tooling component is itself a methodological caution: even the
oracle requires auditing before a shortfall is attributed to a model.

\subsection{Cross-vendor replication: the Mistral panel}
\label{sec:crossvendor}

The grid above is Claude-only, which leaves one component of the circularity
limitation (\S\ref{sec:limitations}) untested: the code was written by
Claude models and evaluated by Claude models. We therefore re-ran the
identical protocol (same prompts, same source packet, same mechanical
replay, and $N{=}5$) on three Mistral models: \textbf{mistral-large-2512}
(the flagship), \textbf{devstral-2512} (a code specialist), and
\textbf{labs-leanstral-1-5} (a Lean~4 theorem-proving specialist, unrelated
to this study's \emph{bare} \texttt{next()} contract arm).

\begin{table}[ht]
\centering
\caption{Mistral panel: mean pass rate, unconditional (conditional),
$N{=}5$, corpus v2.}
\label{tab:mistral}
\small
\begin{tabular}{lccc}
\toprule
Model & SAM contract & bare \texttt{next()} & \tlaplus{} (constrained) \\
\midrule
mistral-large-2512 & 79.6 & 97.0 & 34.5 (59.1) \\
devstral-2512      & 58.2 & 92.7 & 65.7 (96.6) \\
labs-leanstral-1-5 & 64.6 & 87.7 & 47.5 (57.8) \\
\bottomrule
\end{tabular}
\end{table}

\paragraph{What replicates vendor-independently (seven of seven models).}
The bare contract is best or tied for every model of either vendor, and it
remains the only arm with no uncheckable generations in the entire study:
every bare loss is a genuine semantic miss. Semantic misses are
language-independent within each model. mistral-large misses exactly the
partial-payment guard, the same single window in all five bare generations,
and the same rule appears as incorrect post-states in its \tlaplus{} arm.
devstral's consistent bare misses recur as disabled-action deadlocks in its
\tlaplus{} arm. The same model exhibits the same misreadings regardless of
syntax. On significance at $N{=}5$: bare $>$ \tlaplus{} at the permutation
floor ($p{=}.0079$) for mistral-large ($\Delta{=}{+}.626$) and devstral
($\Delta{=}{+}.269$); leanstral's contrasts are directionally identical but
not significant, since it has the study's highest within-arm variance.

\paragraph{What the panel adds.} The SAM checkability hazard appears at
frontier scale across vendors, with a new signature: mistral-large lost one of
five SAM generations to a structurally dead module, and devstral two. These
were not Haiku's tuple wiring, but hallucinated or misused \emph{reactor}
semantics: an invented \texttt{getProposal()} API whose re-entrant calls
livelock the module, a reactor referencing an undefined variable, and a
reactor that unconditionally auto-advances a state. A different vendor
produces a different hallucination in the same component: the richer
contract surface offers many ways to write plausible but non-executing code.
The \tlaplus{} arm likewise gains a failure mechanism beyond the guard
idiom: mistral-large's three parseable specifications each conjoin
\texttt{txState' = "INITIATING"} with \texttt{UNCHANGED <<txState>>} (with
the comment ``Override above''), which is a declarative conjunction misread
as an imperative assignment with override, permanently disabling the action.

\paragraph{One claim sharpened.} The Claude-panel statement that ``the
\tlaplus{} deficit is a replay-contract artifact, with conditional fidelity
97--100\%'' does \emph{not} extend without modification. It holds where the
losses are idiom and checkability (all Claude models, and devstral at
96.6\%). For mistral-large and leanstral, the corpus's rule-coverage windows
expose genuine semantic misses in \tlaplus{} as well (conditional 59.1\% and
57.8\%), which are the \emph{same} misses as in their JavaScript arms. The
general statement is the language-independence of semantic error, not a
universal conditional ceiling. Specialization provided no benefit: the
Lean~4 prover specialist showed no \tlaplus{} advantage (two of five
specifications were unparseable), and the code specialist is mid-pack.
Absolute level tracks general capability rather than domain tuning.

\paragraph{The replication is of ordering and taxonomy, not of level.}
The panel must not be interpreted as vendor parity. Mistral's flagship does
not reach Claude-frontier reliability on this task: its best cell is 97.0\%,
against Fable~5's 100\% on all three arms (fifteen of fifteen generations
perfect on both corpora), and mistral-large produced one structurally dead
SAM specification and two unparseable \tlaplus{} specifications, failure
classes that no Claude frontier model exhibited. The cross-tier comparison
is more pronounced: Claude's mid-tier Sonnet~4.6 outperforms Mistral's
flagship on every arm (98.8 / 99.4 / 88.3 versus 79.6 / 97.0 / 34.5), so the
between-vendor capability gradient is larger than the between-arm gradient
within either vendor. Moreover, the flagship's characteristic miss is
systematic rather than sampled: mistral-large drops the identical rule, the
partial-approval guard, in five of five bare generations and again in its
\tlaplus{} arm. For practitioners the implication is straightforward: at
this level of difficulty, model choice dominates language choice. In this
study, only Fable~5 was perfect across all three arms, with Opus~4.8 perfect
in the bare arm; we make no stronger claim, since fifteen generations on one
protocol measure this task rather than deployments. The relevant caveats
are: a single protocol, $N{=}5$, a prompt family originally iterated against
Claude models in the companion study (a prompt-fit confound that we cannot
exclude), and the fact that two of the three Mistral models are specialists,
so the flagship comparison is the operative one.

\section{Comparison with the companion study}
\label{sec:crossstudy}

The companion study~\cite{companion} concluded with a deliberately
conditional ranking (contract structure first, prompt second, language not
at all), described as ``conditional on the spinlock factorial until
replicated on a richer task,'' with derivation at depth identified as the
decisive open test. This study was designed to discharge that condition, and
it does so on its central axis. Table~\ref{tab:cross} aligns the two
studies.

\begin{table}[ht]
\centering
\caption{The companion study's findings against this study's, per
mechanism.}
\label{tab:cross}
\small
\begin{tabular}{p{0.26\textwidth}p{0.32\textwidth}p{0.32\textwidth}}
\toprule
Finding & Benchmark (spin, locksvc) & Production (finixpos) \\
\midrule
Bare contract & 100\% every model --- saturated & Best or tied every
model; gradient reopens (95.8--100\%) \\
\tlaplus{} deficit & Guard idiom only; conditional 100\% & Guard idiom
(plus one tooling flake); conditional 97--100\% (Claude) \\
SAM ceremony & Taxes weaker models (Haiku 40.7\%) & Aggravated by
SAM-shaped source: 0\%, five of five \\
Language effect & None once shape matched & None: semantic misses
identical across languages, per model \\
Derivation & Spinlock relation guessable (caveat) & Unguessable protocol;
ranking survives \\
\bottomrule
\end{tabular}
\end{table}

Three conclusions follow. First, \textbf{the contract-structure ranking
replicates} on a production system roughly an order of magnitude more
complex, under derivation, and across two vendors: the minimal total
transition contract is what every model writes best; the heavyweight
contract is what the weakest model cannot write at all; and no deficit
anywhere in either study is attributable to the language once the
specification structure and the scoring paradigm are held fixed. Second,
\textbf{the saturation concern is resolved constructively}: the bare
contract saturated the benchmark tasks, and the concern was that it had
eliminated the benchmark's signal. The production protocol shows that the
signal was not eliminated but moved to task difficulty, which now
discriminates cleanly \emph{within} the minimal contract, precisely where a
capability benchmark should have its variance. Third, \textbf{the two
studies jointly support a practical prescription}: when an LLM writes a
specification, provide a minimal, total transition contract over the
observable state, in whatever language the surrounding toolchain prefers;
direct prompt effort only where the contract under-constrains; keep the
heavyweight machinery (SAM's runtime and TLC's model checking) on the
tooling side of the boundary, where it verifies what the model wrote rather
than burdening the writing; and audit the oracle with the same rigour as the
specifications, since the contracts, the replay paradigms, and the model
checker itself each contributed failure components that could have been
misattributed to model incapacity. What the pair of studies leaves open is
unchanged in kind but smaller in scope: a single oracle paradigm, and no
independent replication as yet.

\section{Discussion}
\label{sec:discussion}

\paragraph{The methodology proves productive on its first application to
production code.} The apparatus built for a benchmark (observable-state
projection, dispatch-as-step trace capture, hand-built environment models,
failure-mode-gated exhaustive checking, and mutation-validated corpora)
found real failure-handling gaps in a deployed payment system, reproduced
three of them as real executions, and verified a fix, within roughly a day
of machine-driven work. A single follow-up probe of the production sandbox
then found an integration defect that the entire emulator-based apparatus
was structurally unable to detect. The cost profile matters more than the
defect count: it is what makes formal verification viable
for the large majority of software that has never had a specification,
provided the code has a projectable transition core.

\paragraph{Pattern-embedded semantics.} This code was SAM-shaped from the
outset: the dispatch is the step, the observable state is a projection of
the model, and extraction of the transition relation was mechanical rather
than interpretive. The specification-versus-code gap that SysMoBench's
conformance phase exists to police was closed by construction at the model
core. The audit effort was directed at the \emph{environment} (the ladders,
the races, and the sweep), which is precisely where all seven gaps occur.
This is the constructive reading of the companion study's thesis,
demonstrated on production code: if the transition semantics are embedded in
the program's structure, verification effort concentrates where the risk
actually lies. The counterpoint is the source-shape aggravation finding: the
same SAM structure that made the code auditable made the SAM generation arm
worse, because models imitate the source idiom rather than the contract
instruction. Semantics embedded in the artifact were verified; ceremony
imitated from the artifact produced non-executing specifications. The
bare-contract design implication of the companion study survives its
production test intact.

\paragraph{Derivation at depth: resolved for sources with an explicit
transition core; the implicit case remains open.} The companion study's
decisive open question was whether the contract-structure ranking survives
systems whose semantics cannot be inferred from action names. This protocol
cannot be so inferred: the two rewrites, the glitch guards, and the outcome
fan-out exist only in the source. Across the full grid, the ranking
survives. However, the same property that made extraction mechanical also
bounds the claim: the source contains a \emph{literal transition table}
together with two override acceptors, so deriving \texttt{next()} from it is
closer to transcription plus two insights than to derivation from
unstructured code. The study's clearest example (a Sonnet generation that
transcribed the table verbatim and missed only the override acceptor)
demonstrates this. The accurate statement of what has been established is
that the derivation question is answered \emph{for sources with an explicit
transition core}. Derivation from code in which the state machine is
implicit (scattered across handlers, flags, and callback timing, which is
the far more common case) remains open and is now the relevant frontier.
This paper claims credit for either extraction ease or derivation difficulty
per component, never both for the same component: the model core was
straightforward for exactly the reason it was auditable, while the
environment ladders, where every gap occurred, are where the difficult
derivation lay.

\section{Limitations}
\label{sec:limitations}

\begin{itemize}
\item \textbf{Statistical resolution is at the permutation floor.} The grid
  is complete ($N{=}5$ per cell), but the generations collapse to few
  distinct behaviours; in accordance with the companion study's analysis of
  record, $\Delta$ is the primary quantity and $p$-values are read
  descriptively. Contrasts involving Fable are ceiling-unidentifiable, since
  a perfect grid supports no effect ordering.
\item \textbf{The correlated oracle is demonstrated, not hypothetical.} The
  ground truth for every trace, every reproduction, and the patch's scenario
  validation is the repository's own emulator, which shares the repository,
  the author-and-model lineage, and the understanding of Finix with the code
  under audit. Where the code and the emulator share a misreading, both
  agree and the audit cannot detect the error; \S\ref{sec:sandbox} shows
  that this occurred (the 422 body shape), and it was found only by testing
  against the production sandbox rather than the correlated pair. The trace
  corpus inherits the same limitation (\S\ref{sec:method}). All
  ``reproduced'' claims are therefore \emph{emulator}-reproduced claims,
  except where the sandbox section states otherwise. Two environment
  assumptions remain flagged: partial authorization (documented and
  production-plausible) and cancel-versus-create ordering (undocumented,
  under a conservative reading).
\item \textbf{The correctness verdict is scoped, not absolute.}
  \S\ref{sec:whatchecked} is the authoritative statement: safety invariants
  only, no fairness, and no liveness beyond safety surrogates; the
  environment model is validated by line citation, three real-execution
  reproductions, and post-fix corpus discrimination, rather than by
  mechanical conformance; and process crash--restart and terminal power loss
  are accepted operational risk, with the boundary examined
  (\S\ref{sec:extmodel}). Multi-terminal and multi-order interleavings,
  clock skew beyond the timeout abstraction, and read reordering deeper than
  one step remain unmodelled.
\item \textbf{The oracle commitments are inherited, not resolved.} The
  totality requirement penalizes enabling-condition \tlaplus{} by
  construction (conditional scores are co-reported), and the functional
  replay is licensed by per-scenario determinism; both are restated from the
  companion study, and both remain open.
\item \textbf{The companion study is an unpublished self-citation.} The
  methodology's provenance cannot yet be independently reviewed. This paper
  mitigates the concern by being self-contained (the contracts in
  Appendix~\ref{app:contracts}, the scoring definitions in
  \S\ref{sec:method}, and the verification envelope in
  \S\ref{sec:whatchecked} are all stated here) and by including the study
  plan, committed before the grid was run, in the artifacts. The
  ``pre-committed'' claims are checkable against that file's git history
  rather than against a published registry.
\item \textbf{The observable projection excludes} tips, split payments, and
  the dashboard record-locally flow.
\item \textbf{Fourfold circularity.} The production code was written by
  Claude models; the auditing models, reference-model builders, and patch
  authors are Claude models; the system and the SAM pattern belong to the
  author; and the system under test is public on GitHub~\cite{kizo}, so
  training-data contamination must be addressed rather than assumed away. For
  the Claude checkpoints it is ruled out by release dates: every evaluated
  model's training data ends by January 2026 (Haiku: July 2025), whereas the
  code was first written in February--March 2026 and first made public on
  2026-03-10. The system under test did not exist at any evaluated cutoff, so
  Fable's perfect grid cannot be memorization. The same dates argument covers
  mistral-large-2512 and devstral-2512 (December 2025 releases). For
  labs-leanstral-1-5, which was released after the repository became public
  and has no stated cutoff, dates cannot rule contamination out, and the
  artifact scan is the operative check. That scan (all 105 generations
  against identifiers from the repository \emph{beyond} the prompt's source
  packet) found no domain-token hits for either vendor. Two residuals remain:
  the models do know the SAM library and pattern from training (both long
  predate the cutoffs), which is a background prior relevant to the SAM-arm
  interpretation; and any future re-run on models trained after mid-2026 must
  treat this task as potentially contaminated. The mitigations for the other
  three components are mechanical (real-execution ground truth, line-cited
  counterexamples, controls identical across languages, and findings that are
  unfavourable to every one of the author's artifacts, including the emulator
  itself), but a reader is entitled to weigh the fact that a model family
  audited code that its own family wrote. Independent replication (different
  models, different auditors, and ideally the SysMoBench maintainers' own
  pipeline) is the appropriate response, and all artifacts are provided for
  that purpose.
\end{itemize}

\section{Open questions}
\label{sec:openquestions}

This paper documents a phenomenon; it does not explain it. It reports,
across two studies, seven models, and two vendors, that when an LLM writes a
specification, the minimal total transition contract is what it writes most
reliably, and that the language carrying that contract does not affect the
result. This is a stable phenomenon, but it is not yet an explanation, and
it should not be read as one. The central question it raises is the one it
cannot answer: \emph{can a specification written in ordinary code match or
exceed one written in \tlaplus{}, and if so, why?}

The trade-off must be stated before the question can be posed accurately.
The bare \texttt{next()} contract achieves its fidelity by discarding
exactly what distinguishes SAM: the pattern itself, the acceptor/reactor
decomposition, the auxiliary state, and the model--view--action separation.
What remains is a pure function over the observable state, so the result
supports only the narrow claim that \emph{the minimal transition contract is
what current models write most reliably}, not the broad claim that \emph{code
is a good specification language}. Two interpretations are consistent with
the data, and this study cannot separate them. Under the constructive
interpretation, code that embeds its transition semantics in its structure
is a genuinely good specification medium, and the \tlaplus{} deficit is a
ceremony cost. Under the deflationary interpretation, the advantage is
illusory: the model succeeds by discarding semantics it cannot master, so
the bare contract's advantage is the \emph{absence of failure surface}
rather than the presence of expressive power. The source-shape aggravation
finding (\S\ref{sec:comparison}), in which the richer SAM surface made a
weaker model strictly worse, favours the deflationary interpretation.

A second confound is structural, and this study is not well placed to
resolve it: every link in the chain that produced the Claude grid is Claude.
The production POS was written by Claude models in SAM semantics; the source
packet that every arm must formalize \emph{is} that Claude-authored code;
the prompt family was iterated against Claude in the companion study; and the
reference models and controls were built by Claude. The Mistral panel
(\S\ref{sec:crossvendor}) was introduced to break one component of this
circularity, and it widened the gap that it was intended to probe: Mistral's
flagship trails Claude's mid-tier on every arm. However, the panel cannot
determine why, and two hypotheses are equally consistent with the results.
Under the \emph{capability} hypothesis, the Claude frontier is simply better
at deriving a transition relation from production source. Under the
\emph{familiarity} hypothesis, a model formalizes code most faithfully when
the code was written in its own family's idiom, so Mistral is measured not on
specification skill but on the difficulty of reading a foreign-authored
artifact, and a Claude model asked to formalize Mistral-authored code might
exhibit a symmetric difficulty. The evidence is mixed. In favour of
capability: the semantic misses are language-independent \emph{within} each
model (mistral-large drops the partial-approval guard in JavaScript and
\tlaplus{} alike), which indicates a comprehension failure rather than an
idiom-friction artifact. In favour of familiarity: the source is
Claude-authored SAM code, the most Claude-specific object in the study, and
the prompts were tuned on Claude. The obvious discriminating experiment has not been run: have a non-Claude model, or a human, write a semantically
equivalent POS, regenerate the source packet from it, and re-run the full
grid. If the Claude advantage survives on foreign-authored source, it is
capability; if it diminishes, this paper measured familiarity. Until that
experiment is run, the accurate reading of the seven-of-seven replication is
that the \emph{ordering} of contracts is vendor-independent, while the
\emph{absolute level} may be a property not of the models but of who wrote
the code they were asked to read.

Separating the two is a research program rather than a single paper. If
code-as-specification can match \tlaplus{}, then several questions arise.
What are the semantics of a specification that is ``just code'', that is,
from what do its meaning and its checkable guarantee derive, if not from a
defined logic? Do those semantics survive translation across languages, from
JavaScript to Python to Rust, or is the result an artifact of JavaScript's
prevalence in the training distribution? Should developers adjust their
coding style to make code auditable by construction, and is that a genuine
engineering discipline or a post-hoc rationalization of what the model
happened to prefer? Should SAM be refactored toward its bare core, or does
refactoring \emph{toward} \texttt{next()} amount to abandoning the pattern
rather than validating it? Underlying all of these is a single question: are
we measuring a property of specifications or a property of current models?

This last question is most acute for SAM, for a reason that the \tlaplus{}
comparison obscures. SAM is nearly absent from the training distribution: a
small number of mostly pedagogical examples exist publicly, but nothing of
the scale or variety from which a model could absorb the pattern by
imitation. \tlaplus{} is the opposite: a textbook canon, curated example
suites, and decades of published specifications, a corpus large enough to
imitate. A model therefore cannot reach SAM competence in the way it reaches
\tlaplus{} competence, by pattern-completion over material it has seen; it
would have to \emph{reason} its way to the pattern's semantics from the
contract alone. The grid shows that, given that choice, the models decline to
do so. They reject SAM's decomposition and revert to the transition-relation
form that extensive \tlaplus{} exposure has made familiar, which is precisely
the \texttt{next()} that the \tlaplus{} and bare arms share. On this reading,
the bare contract's advantage is not evidence that minimal code is the right
medium for a specification; it is evidence that models revert to the idiom
that is best represented in their training data. The concerning possibility
is that the study observes not genuine reasoning about how to specify a
system but imitation of the best-represented specification form, and that a
model that had actually mastered SAM, rather than pattern-matched around it,
might specify this protocol \emph{better} in SAM than in either bare arm,
inverting the ranking. The finding would then be a snapshot of a capability
frontier, namely what current models have seen enough of to imitate, rather
than a general principle about how systems should be specified. Nothing here
can separate reasoning from imitation, because the single variable on which
the two explanations disagree, namely what the models absorbed in training,
is exactly the variable that this study cannot control. This paper is
entitled to raise these questions; it is not in a position to answer them.

\section{Conclusion}
\label{sec:conclusion}

A methodology developed on a benchmark spinlock transferred to a production
payment protocol without structural change. In approximately a day it
produced: a core protocol that is correct relative to a validated model
under a precisely stated failure model (\S\ref{sec:whatchecked}); seven
failure-handling gaps, nearly all with a single root cause, three of them reproduced as
real executions, two of which short-charged real orders in the reproduction
runs; and a 24-hunk fix re-verified by the same pipeline, with every
counterexample eliminated under all failure gates, a re-check that
identified a bookkeeping defect in the patch itself, and a validated
follow-up that closed it. The failure-model extensions strengthened rather
than weakened the result: the crash--restart model identified the durability
cases it was designed to find (documented as accepted operational risk in
this staff-present, small-ticket deployment); the two-attempt model
confirmed the anticipated undetected double-charge composition pre-fix and
localized its post-fix residue to one unguarded entry path; and a single
probe of the production Finix sandbox exposed a response-shape divergence
that renders an entire recovery ladder unreachable. This is the
correlated-oracle problem, observed on the paper's own apparatus, and it is
the strongest argument for validating audits against the production API. In
addition, the paper provides a complete replication of the companion study's
central finding (contract structure, not language, governs what LLMs specify
reliably), with two new observations: a specification contract can be
undermined by the very source idiom that it asks the model to formalize, and
task difficulty, not contract weight, is what restores a benchmark's
discriminating power once a minimal contract has saturated it. The open
items, in increasing order of what they would establish, are a device-path
(card-present) validation against the production API, and an independent
replication.

\section*{Reproducibility}
All artifacts are located in the SysMoBench fork alongside the companion
study's: the extraction and protocol map (\texttt{tla\_eval/tasks/finixpos/},
verbatim sources in \texttt{source/}), the reference and environment models
with all TLC configurations and annotated counterexamples
(\texttt{reference/}), the instrumentation patch and harness
(\texttt{data/patches/finixpos\_trace.patch}), the corpus
(\texttt{data/sys\_traces/finixpos*/}; coverage table in
\texttt{corpus\_coverage.md}), controls (\texttt{reference/CONTROLS.md}),
the replay and resumable study driver (\texttt{scripts/finixpos\_tv.py},
\texttt{scripts/finixpos\_phase3\_study.py}), raw per-window results and all
saved generations (\texttt{output/finixpos*}), the results document
(\texttt{docs/finixpos\_study\_results.md}), and the fix patches with their
annotated companions and validated previews (\texttt{Merchant/v2/patches/}
in the application repository). The extended-audit artifacts are included
alongside: the sandbox experiment with redacted transcripts
(\texttt{reference/sandbox\_repro/}), the extended failure models and
per-channel enumeration (\texttt{reference/EXTENDED\_FAILURE\_MODEL.md},
\texttt{FinixPOSCrash/Stale/TwoAttempt.tla}; \texttt{FinixPOSRefund.tla}
retired with a header note), the post-fix verification
(\texttt{reference/FinixPOSFixed*.tla},
\texttt{counterexamples/POSTFIX-NOTES.md},
\texttt{data/sys\_traces/finixpos\_fixed/},
\texttt{output/finixpos\_postfix*}), the contamination analysis
(\texttt{tla\_eval/tasks/finixpos/CONTAMINATION.md}), and the cross-vendor
Mistral panel (45 additional generations, rescores, and the seven-model
analysis in \texttt{scripts/finixpos\_final\_analysis.py} /
\texttt{output/finixpos\_final\_analysis.json}). The production source is
open source (the \texttt{kizo-food} merchant application~\cite{kizo}); the
five extracted files are additionally included verbatim in the study corpus
with recorded provenance.

\section*{Acknowledgements}
The author thanks Qian Cheng (PhD student, Nanjing University) and Tianyin Xu
(Associate Professor, University of Illinois Urbana-Champaign) for stimulating
discussions and for being open to the idea of testing the SAM pattern, in its
JavaScript form, as a specification language.

\appendix

\section{The three specification contracts, abridged}
\label{app:contracts}

The full prompts are included in
\texttt{tla\_eval/tasks/finixpos/prompts/}. All three arms receive the
identical source packet, the identical seven-field observable contract, and
the identical action alphabet with data schemas; they differ only in the
module shape the model must produce. The excerpts below are the operative
contract clauses, verbatim.

\paragraph{Bare \texttt{next()} arm --- the transition function.}
\begin{lstlisting}[style=code]
Convert the following production TypeScript source [...] into a plain
JavaScript transition function -- NO libraries, NO frameworks, NO SAM
pattern. Just a pure function over the observable state.

Your specification is a CommonJS module [...]. It MUST export exactly:

    module.exports = { init, next };

  - init() returns the initial state: { txState: 'IDLE', orderId: null,
    amountCents: null, transferId: null, declineCode: null,
    approvedAmountCents: null, paymentId: null }.
  - next(state, action, data) is a PURE function returning the NEW state
    (same seven keys) after applying ONE action.
\end{lstlisting}

\paragraph{SAM arm --- full library contract.}
\begin{lstlisting}[style=code]
Convert the following production TypeScript source [...] into an
executable JavaScript SAM specification using the
@cognitive-fab/sam-pattern library.

module.exports = { instance, init, actions, getState, setState,
                   checkerIntents };

- instance -- the SAM instance, created with:
  createInstance({ instanceName: 'finixpos', hasAsyncActions: false });
- init() -- resets the model to its initial state [...]
- actions -- an object with EXACTLY these mandatory action names and
  data shapes [...] Each is a function that fires the corresponding SAM
  intent synchronously.
- getState() -- a plain JSON-serializable snapshot with exactly the
  seven observable keys (no functions, no SAM internals).
- setState(snapshot) -- forces the model to a given snapshot [...]
\end{lstlisting}

\paragraph{\tlaplus{} arm --- constrained module.}
\begin{lstlisting}[style=code]
Generate a TLA+ specification of the payment workflow below [...]
constrained to the same observable-state contract [as the JavaScript
arms].

Your module MUST be named `finixpos` and its observable state MUST be
EXACTLY these seven variables -- declare no others:

    VARIABLES txState, orderId, amountCents, transferId, declineCode,
              approvedAmountCents, paymentId

Sentinels are plain values -- DO NOT define any value with CHOOSE [...]

    NONE  == "none"
    NOAMT == -1

Declare as CONSTANTS the finite value sets: OrderIds, Amounts,
TransferIds, DeclineCodes, PaymentIds. The harness binds them to small
finite sets that include every value appearing in the traces.

Do NOT introduce any other VARIABLES (no pc, no message channels, no
history). [Action operators: mandatory top-level named signatures.]
\end{lstlisting}

The replay drivers are symmetric: the JS arms load the module and call
\texttt{next} (or fire the SAM intent) per window; the \tlaplus{} arm
synthesizes a per-window harness (\texttt{TVInit} pins $\mathit{pre}$; one
step of the window's action; an \texttt{INVARIANT} asserting the expected
$\mathit{post}$ is \emph{not} reached, so a counterexample is a pass) with
constants bound to the values observed in that window. Statuses are pass,
fail, or unscoreable, as defined in \S\ref{sec:method}.

\end{document}